\pdfoutput=1

\documentclass[twocolumn,prd,preprintnumbers,amsmath,amssymb,superscriptaddress,nofootinbib]{revtex4}

\usepackage{epsfig,psfrag,graphics,verbatim}
\usepackage{graphicx}
\usepackage{dcolumn}
\usepackage{bm}
\usepackage{color}

\newcommand{\be}{\begin{equation}}
\newcommand{\ee}{\end{equation}}

\begin{document}

\title{Dark Matter distribution in the Milky Way: microlensing and dynamical constraints}

\author{Fabio Iocco}
\affiliation{Institut d'Astrophysique de Paris, UMR 7095-CNRS, Univ.~Pierre \& Marie Curie, 98bis Bd Arago 75014 Paris, France}

\author{Miguel Pato}
\affiliation{Institute for Theoretical Physics, University of Z\"urich, Winterthurerstrasse 190, 8057 Z\"urich CH}

\author{Gianfranco Bertone}
\affiliation{Institut d'Astrophysique de Paris, UMR 7095-CNRS, Univ.~Pierre \& Marie Curie, 98bis Bd Arago 75014 Paris, France}
\affiliation{Institute for Theoretical Physics, University of Z\"urich, Winterthurerstrasse 190, 8057 Z\"urich CH}

\author{Philippe Jetzer}
\affiliation{Institute for Theoretical Physics, University of Z\"urich, Winterthurerstrasse 190, 8057 Z\"urich CH}

\date{\today}

\begin{abstract}
We show that current microlensing and dynamical observations of the Galaxy permit to set interesting constraints on the Dark Matter local density and profile slope towards the galactic centre. Assuming state-of-the-art models for the distribution of baryons in the Galaxy, we find that the most commonly discussed Dark Matter profiles ({\it viz.}~Navarro-Frenk-White and Einasto) are consistent with microlensing and dynamical observations, while extreme adiabatically compressed profiles are robustly ruled out. When a baryonic model that also includes a description of the gas is adopted, our analysis provides a determination of the local Dark Matter density, $\rho_0=0.20-0.56\textrm{ GeV/cm}^3$ at 1$\sigma$, that is found to be compatible with estimates in the literature based on different techniques. 
\end{abstract}

\maketitle

\section{Introduction} 
\par The presence of large amounts of Dark Matter (DM) in the Universe, and in particular in the Milky Way, is established on sound observational grounds \cite{BergstromBook,BertoneReview,BertoneBook}. However, a detailed description of the Dark Matter distribution in the Galaxy is hard to achieve, despite the tremendous progress in numerical simulations of galaxy-sized objects over the last few years \cite{VL2,VL2site,Aq,Aqsite}, therefore precluding, among other things, a precise interpretation of direct and indirect Dark Matter searches (see e.g.~Refs.~\cite{StrigariTrotta,Green:2010ri,McCabe:2010zh,paperastrodirect,Vogelsberger,Pieri:2009je}). In order to constrain Milky Way mass models, several observables have been used in the literature, including star counts, the motion of gas and stars or microlensing events. Here, we focus on two key probes, {\it viz.}~microlensing observations and dynamical measurements, and show that interesting constraints can be set on the DM distribution using those measurements only. This provides complementary evidence for the existence of Dark Matter in the Galaxy, with a distribution compatible with that inferred from other observables. 

\par Gravitational microlensing has long been adopted as a tool to study the structure of the Galaxy \cite{Paczynski85,KiragaPaczynski}. This technique was suggested in \cite{Paczynski85} to probe the existence of Massive Compact Halo Objects (MACHOs) by using source stars in the Magellanic Clouds, M31 and M33. Soon after the observation of the first microlensing events in the early 90s \cite{Alcock:1993eu,Aubourg:1993wb,Udalski:1993zz}, it became apparent that MACHOs could not be a dominant component of the galactic dark halo \cite{Gates:1994jy} -- therefore a type of dark, non-compact matter was needed. Meanwhile, microlensing of stars towards the galactic centre proved useful in tuning the existent bulge, bar and disk models. Until today, several thousands of microlensing events have been collected -- most notably by OGLE \cite{Udalski:1993zz,Udalski:1994ei,Sumi:2005da}, MACHO \cite{Alcock:1993eu,Alcock:2000rv,Popowski:2004uv}, EROS \cite{Aubourg:1993wb,Afonso:2003aw,Hamadache:2006fw,2009A&A...500.1027R} and MOA \cite{Sumi:2002wg} campaigns (see \cite{Moniez:2010zt,CalchiNovati:2009af} for a review) -- allowing for precise estimates of the microlensing optical depth towards the galactic bulge, spiral arms and Magellanic Clouds. 

\par Another well-known tool to constrain the different components of our Galaxy is the rotation curve. Unlike microlensing -- which is sensitive to the distribution of compact matter along the line of sight --, dynamical measurements constrain the total mass distribution. Several decades ago, the observation of flat rotation curves in external spiral galaxies provided convincing evidence for Dark Matter, but actually the rotation curve of the Milky Way (a spiral itself) is far from being precisely determined given our peculiar position within the galactic disk. In fact, although hosts of data have been gathered over the years \cite{Sofue:2008wt}, allowing for ever more accurate dynamical models, sizeable uncertainties remain on Milky Way mass models, a circumstance that in turn affects our capability to constrain the DM distribution, as we shall see below.

\par In a pioneering work \cite{Kuijken}, Kuijken pointed out that a non-zero microlensing optical depth towards the galactic bulge sets a lower limit on the enclosed baryonic mass and thus a lower limit on the corresponding circular velocity within the solar circle. Using this result, Binney \& Evans \cite{BinneyEvans} derived an upper limit on the dark matter contribution to the rotation curve, which led them to the conclusion that ``the cuspy haloes favoured by the Cold Dark Matter cosmology (and its variants) are inconsistent with the observational data on the Galaxy'' \cite{BinneyEvans}. This claim was however based on a preliminary determination of the microlensing optical depth by the MACHO collaboration \cite{Popowski:2000jq} (later replaced by a less constraining measurement from the same collaboration \cite{Popowski:2004uv}) as well as a very simplified treatment of the galactic rotation curve. 

\par Here, we revisit this claim in light of recent microlensing observations towards a variety of galactic regions and up-to-date measurements of our Galaxy's rotation curve. A very significant improvement of our work with respect to the existing literature is a proper inclusion of all experimental uncertainties regarding both microlensing and dynamical data, which makes our conclusions sound on statistical grounds. Furthermore, we shall also use a wide range of state-of-the-art models for the galactic baryonic component, so that modelling uncertainties are appropriately bracketed. Our ultimate aim is to combine microlensing and dynamical observables to draw robust constraints on the Dark Matter distribution in the inner region of the Milky Way.

\section{Milky Way modelling}\label{MWmodel}

\par In order to extract useful constraints on the DM distribution from microlensing and dynamical observables, we need to specify a mass model for our Galaxy. In this work, we shall implement a bulge/bar region, a disk and a Dark Matter halo. The Sun sits on the galactic disk at $(x_\odot,y_\odot,z_\odot)=(R_0,0,0)$, where $R_0$ is the galactocentric distance\footnote{Actually, the Sun lies somewhat off the central plane at $z_\odot\lesssim 30$ pc \cite{Majaess:2009xc}, but this is a very small displacement compared to other quantities at play and thus we shall take the reasonable assumption $z_\odot=0$.}. We consider the range $R_0=8.0\pm 0.5$ kpc throughout the analysis, taking 8.0 kpc as our fiducial value. This encompasses recent \cite{Gillessen:2008qv} and less recent \cite{Reid:1993fx} determininations as well as the 1985 IAU recommended values \cite{IAU1985b}. All scale distances in this Section are rescaled as $R_0/8\textrm{ kpc}$.

\par The bulge at the centre of the Milky Way has long been suspected to be bar-shaped. This was confirmed by near-infrared observations \cite{Dwek:1995xu} and the distribution of red clump giant stars \cite{Stanek:1995ws,Rattenbury:2007ky}, which produce convincing evidence for a triaxial bulge with its near end at positive galactic longitudes $\ell>0$ and its major axis inclined $\alpha_b\sim 25^\circ$ with respect to the galactic centre line of sight. Several families of triaxial mass density distributions have been tested against the data, and two parameterisations appear to work particularly well \cite{Dwek:1995xu,Stanek:1995ws}:
\begin{eqnarray}
&\rho_{b}(x,y,z) = \rho_{0,b} \, e^{- r_1}  \quad & \textrm{(E2)} \\
&\rho_{b}(x,y,z) = \rho_{0,b} \, e^{-\frac{r_2^2}{2}}  \quad & \textrm{(G2)} \quad ,
\end{eqnarray}
with 
\begin{equation*}
r_1^2  =  \frac{x'^2}{x_b^2} + \frac{y'^2}{y_b^2} + \frac{z'^2}{z_b^2}  \, , \,
 r_2^4=  \left( \frac{x'^2}{x_b^2} +\frac{y'^2}{y_b^2} \right)^2  + \frac{z'^4}{z_b^4} \quad ,
\end{equation*}
where $(x',y',z')$ are the coordinates along the major, intermediate and minor axes, respectively, and the parameters of these fitting formulae are \cite{Stanek:1995ws} 
\begin{align*} 
 \textrm{E2:}\, \alpha_b=23.8^\circ;(x_b,y_b,z_b)=&(0.899,0.386,0.250)\frac{R_0}{\textrm{\small 8 kpc}} \textrm{\small kpc} \\ 
 \textrm{G2:}\, \alpha_b=24.9^\circ;(x_b,y_b,z_b)=&(1.239,0.609,0.438) \frac{R_0}{\textrm{\small 8 kpc}} \textrm{\small kpc} \, .  
\end{align*}
For both the E2 and G2 models, we implement an exponential cutoff proportional to $\exp(-(R'-R_{m})^2/2r_0^2)$ (with $R'^2=x'^2+y'^2$, $r_0=0.5$ kpc) beyond the co-rotation radius $R_{m}=3.5$ kpc as in \cite{Dwek:1995xu,Novatin:2007dd}. Another relevant bulge model is the one introduced by Zhao in \cite{Zhao:1995qh},
\begin{equation}
\rho_{b}(x,y,z) = \rho_{0,b} \left( s_a^{-1.85} e ^{-s_a}  +  e^{-\frac{r_2^2}{2}} \right) \quad \textrm{(Zhao)} \quad ,
\end{equation}
with $s_a^2=(q_b^2(x^2+y^2)+z^2)/z_b^2$, which is particularly suited to produce a steep rise in the rotation curve at $r\lesssim 0.5$ kpc. Inspired by \cite{CatenaUllio} we shall consider an axisymmetrised version of this profile with $x_b=y_b=0.9\textrm{ kpc}(R_0/8\textrm{ kpc})$, $z_b=0.4\textrm{ kpc}(R_0/8\textrm{ kpc})$ and $q_b=0.6$. Finally, the recent and sophisticated model of Gardner et al \cite{Gardner:2010sa,Gardner:2010pd} is also implemented -- in this case there are separate bar and bulge components:
\begin{eqnarray}
\rho_{b}(x,y,z) = f_{0,b} \left( \rho_{bar}(x,y,z) + \rho_{bulge}(x,y,z) \right) \\ \nonumber \textrm{(Gardner et al)} \quad ,
\end{eqnarray}
where the bar density is given by a Ferrers model \cite{Pfenniger84}
\begin{equation*}
\rho_{bar}(x,y,z) = \left\{\begin{array}{ll}
\rho_{0,bar}\left(1-r_1^2\right)^n & \textrm{for }r_1<1 \\
0 & \textrm{for }r_1\geq 1
\end{array} \right. \quad ,
\end{equation*}
with $n=2$, $\alpha_b=25^{\circ}$, $x_b=3.5\textrm{ kpc}(R_0/8\textrm{ kpc})$, $y_b=1.4\textrm{ kpc}(R_0/8\textrm{ kpc})$, $z_b=1.0\textrm{ kpc}(R_0/8\textrm{ kpc})$, $M_{bar}=10^{10}\textrm{ M}_\odot$ \cite{Gardner:2010sa}, and the bulge component is given in terms of the gravitational potential $\phi_{bulge}$ defined in \cite{Gardner:2010pd}. The mass density associated to $\phi_{bulge}$ is readily obtained using the Poisson equation. The constant $f_{0,b}$ is just a normalisation and its role will become apparent in Section \ref{microl}.

\par As for the stellar disk in our Galaxy, it is well-known that there are a thin and a thick components corresponding to distinct star populations. Instead of covering a whole range of different vertical and radial disk profiles, we adopt two extreme cases: a thin-only disk and a thin+thick disk. Although the former model is somewhat unrealistic, it will prove useful in exploring the present uncertainty associated to disk modelling. For an alternative disk model see Ref.~\cite{Novati:2011ii}. Following \cite{Novatin:2007dd,Han:2003ws},
\begin{eqnarray}
\rho_{d}\left(R,z\right) = \frac{\rho_{0,d}}{\eta} e^{-\frac{R-R_0}{H}} \left[(1-\beta)\textrm{sech}^2\frac{z}{\eta h_1} + \beta e^{-\frac{|z|}{\eta h_2}} \right] \\ \nonumber \textrm{(Han \& Gould)} 
\end{eqnarray}
with 
\begin{align*}
R^2=x^2+y^2\, , \, &H=2.75\frac{R_0}{8\textrm{ kpc}} \textrm{kpc} \, , \\ 
h_1=0.270 \frac{R_0}{8\textrm{ kpc}}  \textrm{kpc} \, &, \, h_2=0.440\frac{R_0}{8\textrm{ kpc}} \textrm{kpc} \, , \\
\eta(R)=\max \Big[ 0.670&, 0.114+ \frac{R}{9.025\textrm{ kpc}} \Big] \\
\textrm{thin: } \beta= 0 \, , \, \rho_{0,d}=&4.4\times 10^7 \eta(R_0) \textrm{ M}_\odot \textrm{/kpc}^{3}  \\
\textrm{thin+thick: }  \beta=0.565& \, , \, \rho_{0,d}=4.93\times 10^7 \textrm{ M}_\odot \textrm{/kpc}^{3} \, .
\end{align*}
Again, we shall also refer to the model of Gardner et al where both stellar and gas disks are modelled through the corresponding gravitational potentials (see details in \cite{Gardner:2010pd}).

\par In the remainder of the work, five models for the galactic baryonic component (i.e.~bulge/bar and disk) will be used:
\begin{itemize}
\item {\bf Model 1:} E2 bulge and thin+thick disk;
\item {\bf Model 2:} G2 bulge and thin+thick disk;
\item {\bf Model 3:} G2 bulge and thin disk;
\item {\bf Model 4:} Zhao bulge and thin disk; and
\item {\bf Model 5:} Gardner et al bulge/bar, stellar disk and gas disk.
\end{itemize}
We emphasise that these fiducial setups reasonably bracket the uncertainties in modelling the galactic baryonic component, and thus will be useful in assessing how our results depend on such modelling.

\par The last piece missing in our Milky Way mass model is the Dark Matter halo. In view of the findings of numerical simulations \cite{VL2,Aq}, we implement spherically symmetric generalised Navarro-Frenk-White (NFW) and Einasto profiles:
\begin{eqnarray}\label{NFW}
\rho_{DM}(r)=&\bar{\rho}_s (r/r_s)^{-\alpha}(1+r/r_s)^{-3+\alpha} \, \, &\textrm{(NFW)} \, \\ \label{Ein}
\rho_{DM}(r)=&\bar{\rho}_s \exp\left[-\frac{2}{\alpha} \left( \left(\frac{r}{r_s}\right)^\alpha-1 \right)\right] \, \, &\textrm{(Einasto)} ,
\end{eqnarray}
where $r^2=x^2+y^2+z^2$, $r_s$ is the scale radius, $\bar{\rho}_s$ is the scale density and $\alpha$ is the inner slope for the NFW profile and a shape parameter for the Einasto profile. Typical ranges found in $N$-body simulations are $0.9\lesssim \alpha \lesssim 1.2$ in the case of NFW \cite{VL2,Navarro:2008kc} and $0.12\lesssim \alpha \lesssim 0.22$ in the case of Einasto \cite{Navarro:2003ew,Hayashi:2007uk,Gao:2007gh,Navarro:2008kc}. In the following, the normalisation of the DM profile will be set by the local DM density, $\rho_0\equiv\rho_{DM}(R_0)$, that can be easily cast in terms of $\bar{\rho}_s$. We shall take a fiducial interval $\rho_0=0.4\pm0.1\textrm{ GeV/cm}^3$ in line with recent determinations \cite{CatenaUllio,SaluccilocalDM,paperDMlocal,Garbari:2011dh}.

\par Regarding the scale radius $r_s$, DM-only simulations tell us that the virial concentration $c_{vir}=r_{vir}/r_s$ of an object of mass $M_{vir}=10^{12}h^{-1}\textrm{ M}_\odot$ lies in the range $\log_{10} c_{vir}=0.9-1.1$ (at 1$\sigma$) \cite{Maccio':2008xb} which translates into $r_s\sim 24-38$ kpc. However, works where galactic dynamical observables are used to constrain a Milky Way mass model \cite{CatenaUllio} (see also \cite{UllioBuckley}) seem to favour higher concentrations, $c_{vir}=10-25$, or $r_s\sim 12-30$ kpc. Hence, we shall take in the following a rather wide range $r_s=20^{+15}_{-10}$ kpc. 

\par We stress that the Dark Matter profiles found in numerical simulations deviate significantly from spherical symmetry: while DM-only simulations lead to very prolate shapes, the inclusion of baryons leads to more oblate (but still triaxial) DM distributions \cite{Debattista2007,ATM10}. In order to check the relevance of a non-spherical DM halo for our purposes, we shall also consider the NFW profile in equation \eqref{NFW} with $r$ replaced by $m=(x^2+y^2+z^2/q^2)^{1/2}$ and axis ratio $q=0.7$. This represents an oblate profile with a shape compatible with the results of \cite{paperDMlocal}.

\par As of today, it is not clear how baryons affect the DM profile in our Galaxy. One possibility is that when the Galaxy formed and the baryons contracted towards the centre, the dissipationless component was dragged leading to a steepening of the DM profile. This was proposed long ago in Ref.~\cite{Blumenthal}, where a simple adiabatic contraction model was introduced (more refined models have also been constructed, see Refs.~\cite{Gnedin2004,Gustafsson2006}): given a spherically symmetric initial total mass distribution $M_i(<r_i)$ and a final baryon distribution $M_b(<r)$, the final DM distribution $M_{DM}(<r)$ obeys \cite{Blumenthal}
\begin{equation}\label{AC}
r\left(M_b(<r)+M_{DM}(<r)\right) = r_i M_i(<r_i) \quad ,
\end{equation}
with $M_{DM}(<r)=(1-f_b)M_i(<r_i)$, $f_b$ being the total baryonic fraction in the Galaxy. Taking $M_i(<r_i)$ to be the initial mass distribution associated to the NFW profile \eqref{NFW} and $M_b(<r)$ the baryonic mass distribution given by models 1--5, we can easily solve \eqref{AC} for $r$ while fixing $r_i$ and $f_b$. This sets the final mass distribution $M_{DM}(<r)$ and (upon derivation) the DM profile. For the sake of completeness, in our analysis we shall also use the described procedure to model the DM profile and therefore test adiabatic contraction models against galactic dynamical observables.

\par Finally, note that in the following, given a baryonic model and assumed $r_s$, we let vary $\rho_0$ and $\alpha$. These will be the main phenomenological parameters in our analysis.

\section{Microlensing}\label{microl}
\par In this Section we present a brief overview of the gravitational microlensing formalism and of the most recent observations.

\subsection{Theoretical framework} 
\par Gravitational microlensing is a direct consequence of Einstein's General Relativity: as a massive compact object -- the lens -- moves across the line of sight towards a given luminous source, the source light gets deflected and two distinct images form, assuming a Schwarzschild lens. In the microlensing regime these images are not resolved and hence the observable effect is a temporary magnification of the source. For a point-like source and a point-like lens, the magnification of the source reads \cite{Paczynski85}
\begin{equation}
\label{magnification}
A(t)=\frac{u(t)^2+2}{u(t)\sqrt{u(t)^2+4}} \quad ,
\end{equation}
where $u(t)$ is the transverse distance between the lens and the line of sight measured in units of the Einstein radius $R_E=\sqrt{(4GM_l/c^2) D_l(1-D_l/D_s)}$. Here, $D_l$ ($D_s$) is the distance between the observer and the lens (source) and $M_l$ is the lens mass. The typical duration of a microlensing event is set by the Einstein radius and the lens velocity in the plane perpendicular to the line of sight $v_T$: $t_E=R_E/v_T$. For galactic lens objects with masses ranging from $10^{-6}$ M$_{\odot}$ to $10^{2}$ M$_{\odot}$, the typical microlensing time scale varies from a few hours to a few years \cite{Paczynski85}. Therefore, a thorough monitoring of galactic stars in different directions can be used to study a whole population of low-luminosity objects along the line of sight, which would otherwise remain undetected.

\par The probability of observing the microlensing of a given luminous source is driven by the mass distribution of lenses $\rho_l$ along the line of sight. A particularly interesting quantity is the optical depth $\tau$ which quantifies the probability that a lens comes within one Einstein radius of the line of sight, or equivalently that the magnification of the source exceeds $A=1.34$ (cf.~equation \eqref{magnification} with $u=1$). If all sources towards galactic coordinates $(\ell,b)$ are placed at the same distance $D_s$ (we will relax this simplifying assumption for a more general formulation later on), the theoretically expected optical depth can be written as:
\begin{equation}
\label{OD}
\tau(\ell,b,D_s)=\frac{4\pi G}{c^2}\int_0^{D_s} dD_l \,\, \rho_l(\ell,b,D_l) \, D_l \left(1-\frac{D_l}{D_s}\right) \quad ,
\end{equation}
where $\rho_l$ refers to all possible lens objects (in our case the bulge/bar and the stellar disk). For our purposes it is important to stress that {\it (i)} microlensing is caused only by massive compact bodies, and not by gas nor Dark Matter; and {\it (ii)} the optical depth $\tau$ is independent of the lens mass function, since the surface of the Einstein disk is proportional to $M_l$. As a consequence, measurements of the optical depth depend only on the line of sight integral of the mass density of lenses $\rho_l$ as emphasised in equation \eqref{OD}, and will therefore set the normalisation of our baryonic models. 

\par It is to be noted that equation \eqref{OD} is obtained under the assumption that all the sources lie at the same distance $D_s$. This is no longer valid for microlensing observations towards the galactic centre, as self-lensing (caused by the fact that a ``source'' in a lensing event is also acting as lens for another event) plays an important role. In this case one has to integrate not only on the distance of the lenses but also on the distance of the sources, since the number density of both sources and lenses can vary substantially \cite{KiragaPaczynski}:
\begin{equation}
\label{weightedOD}
\langle\tau\rangle(\ell,b)= \frac{\int_0^{r_{\infty}} dD_s \, \, \tau(\ell,b,D_s) \, dn_s/dD_s } { \int_0^{r_{\infty}} dD_s \, \, dn_s/dD_s   } \quad ,
\end{equation}
where $dn_s/dD_s\propto \rho_s(D_s) D_s^{2+2\beta_s}$ is the distance distribution of detectable sources, $\rho_s$ is the mass density of sources and $r_{\infty}$ is the maximum distance at which sources can be found that we fix to $r_\infty=20$ kpc (our results do not depend much on the specific choice of this parameter). Notice that the expression for $dn_s/dD_s$, introduced in \cite{KiragaPaczynski}, includes both a volume effect in the $D_s^2$ factor since more sources lie at greater distances, and a luminosity effect in the factor $D_s^{2\beta_s}$ given that only sources above the threshold luminosity can be detected (assuming that the distribution of sources above luminosity $L$ follows $L^{\beta_s}$). The index $\beta_s$ depends, of course, on the type of source and it is usually taken to be $\beta_s\lesssim -1$ for main sequence stars and $\beta_s=0$ for red giant clump stars \cite{Zhao:1996ms,Han:2003ws}. In the following Section we explain how we use $\rho_s$ and $\rho_l$ to recover the expected optical depth from each of the  models defined in Section \ref{MWmodel}, while comparing it with observations.

\subsection{Observations}

\par Early measurements of the optical depth towards the galactic bulge \cite{Alcock:2000rv,Sumi:2002wg,Popowski:2000jq} showed values significantly higher than what expected from initial estimates \cite{Paczynski85,KiragaPaczynski}. This motivated the study in \cite{BinneyEvans}, as previously discussed, in the then-justified belief that the higher density of lenses for microlensing would leave less room for the gas and Dark Matter components in our Galaxy. However, since those early measurements, huge progress has been done in microlensing observations and modelling of the bulge (see Section \ref{MWmodel} for this latter point): in fact, the MACHO, EROS and OGLE collaborations have recently performed measurements of the optical depth towards the galactic centre. The present values are less constraining than the early results, thus making it compelling to perform a more accurate analysis of the different mass components of our Galaxy.

\begin{figure}
\centering
\includegraphics[width=0.45\textwidth]{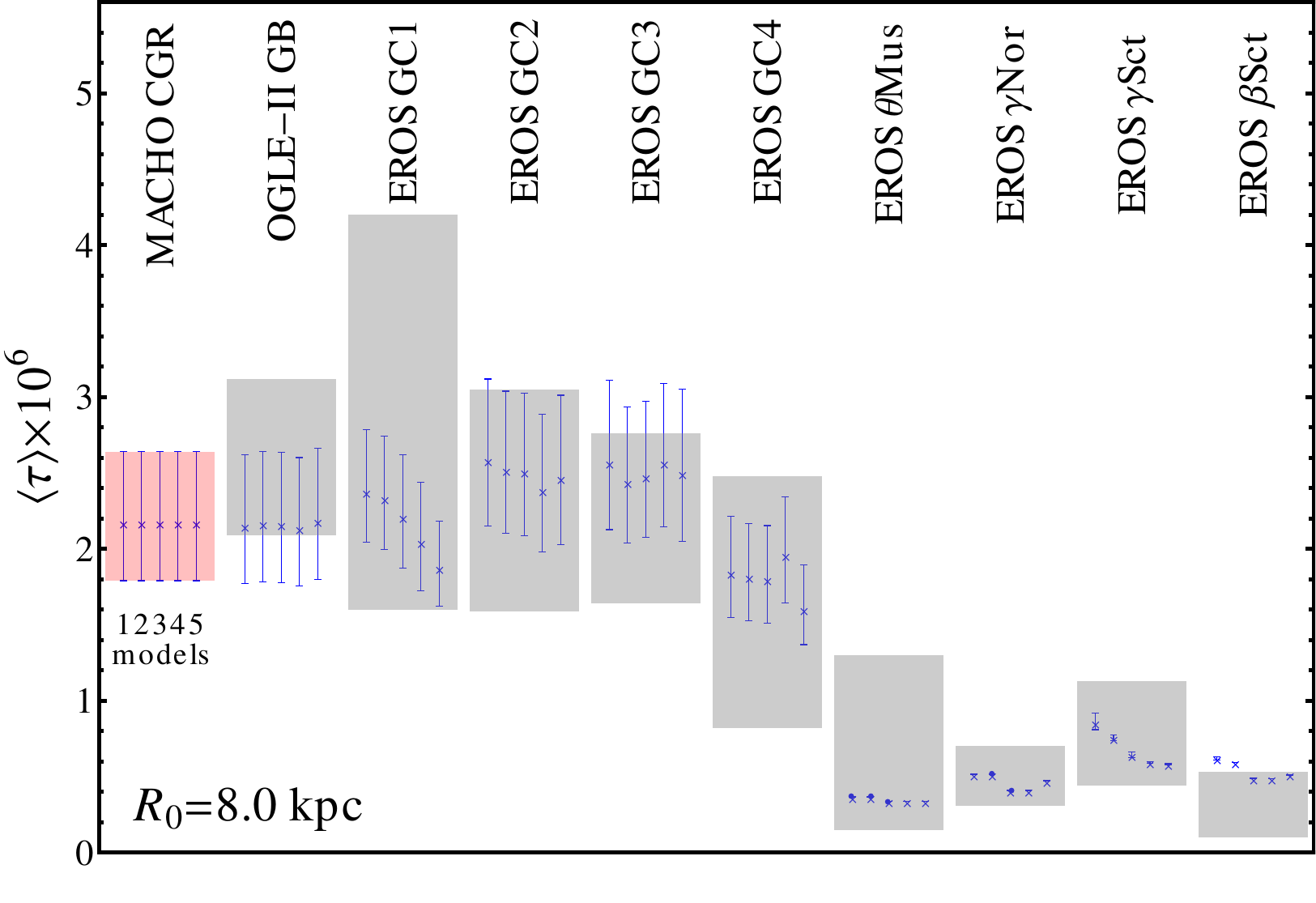}
\caption{Measured and expected microlensing optical depths for different galactic regions. The measured optical depths along with the corresponding uncertainties are shown by the boxes, while the expected values from our baryonic models are shown by the blue dots and error bars. All five models defined in Section \ref{MWmodel} have been scaled to agree with the MACHO \cite{Popowski:2004uv} measurement towards $(\ell,b)=(1.5^\circ,-2.68^\circ)$ represented by the leftmost box. Using this normalisation we show the expected optical depth for different fields of view observed by several campaigns: OGLE-II GB \cite{Sumi:2005da}, EROS GC1-4 \cite{Hamadache:2006fw} and EROS $\theta$Mus, $\gamma$Nor, $\gamma$Sct, $\beta$Sct \cite{2009A&A...500.1027R}. In this plot all uncertainties are 1$\sigma$ and $R_0=8$ kpc.} 
\label{fig:depth}
\end{figure}

\par As a default, we start by adopting the 2005 measurement of the MACHO collaboration towards the galactic bulge \cite{Popowski:2004uv}. Out of the 94 observed fields, they identify 9 at very small galactic latitude and longitude as the Central Galactic Region (CGR), and find the average value 
\begin{equation}\label{macho}
\langle\tau\rangle=2.17^{+0.47}_{-0.38}\times 10^{-6}\textrm{ for }(\ell,b)=(1.50^\circ,-2.68^\circ) \quad .
\end{equation}
We scale our bulge models to meet this optical depth measurement as done in \cite{Novatin:2007dd}: assigned all model parameters but $\rho_{0,b}$ for models 1--4 or $f_{0,b}$ for model 5, we set the latter by requiring that the optical depth obtained using equation \eqref{weightedOD} matches the value in equation \eqref{macho}. We checked that our optical depth results for models 1 and 2 are in agreement with those obtained in Figure 1 of Ref.~\cite{Novatin:2007dd}. Since the source stars monitored in \cite{Popowski:2004uv} are red clump giants in the galactic bulge, we fix $\beta_s=0$ in equation \eqref{weightedOD} while using $\rho_s\equiv \rho_b$ and $\rho_l\equiv \rho_b+\rho_d$. Notice that this procedure gives an $1\sigma$ range for $\rho_{0,b}$ or $f_{0,b}$ corresponding to the experimental uncertainty in equation \eqref{macho}. 

\par We repeat the procedure for each model discussed in Section \ref{MWmodel}, thus fixing the galactic baryonic component. It is now possible to cross-check our models against current observations from other experimental groups. We do so by computing the optical depth through equation \eqref{weightedOD} making use of the just-defined bulge normalisations. In particular, we consider the following measurements:
\begin{itemize}

\item OGLE-II GB \cite{Sumi:2005da}: $\langle\tau\rangle=2.55^{+0.57}_{-0.46}\times 10^{-6}$ for $(\ell,b)=(1.16^\circ,-2.75^\circ)$. In this case we use $\beta_s=0$, $\rho_s\equiv \rho_b$ and $\rho_l\equiv \rho_b+\rho_d$ in equation \eqref{weightedOD} since the source stars are red clump giants in the galactic bulge.

\item EROS GC1--4 \cite{Hamadache:2006fw}: 
$\langle\tau\rangle=2.90\pm1.30 \times 10^{-6}$ for $(\ell,b)=(-4.5^\circ,2.40^\circ)$;
$\langle\tau\rangle=2.32\pm1.73 \times 10^{-6}$ for $(\ell,b)=(-1.5^\circ,2.42^\circ)$;
$\langle\tau\rangle=2.20\pm1.56 \times 10^{-6}$ for $(\ell,b)=(1.5^\circ,2.22^\circ)$;
$\langle\tau\rangle=1.65\pm0.83 \times 10^{-6}$ for $(\ell,b)=(4.5^\circ,2.53^\circ)$.
In this case we use $\beta_s=0$, $\rho_s\equiv \rho_b$ and $\rho_l\equiv \rho_b+\rho_d$ in equation \eqref{weightedOD} since the source stars are red clump giants in the galactic bulge.

\item EROS spiral arms \cite{2009A&A...500.1027R}: 
$\langle\tau\rangle=0.67^{+0.63}_{-0.52}\times 10^{-6}$ for $(\ell,b)=(306.56^\circ,-1.46^\circ)$
$\theta$Mus; $\langle\tau\rangle=0.49^{+0.21}_{-0.18}\times 10^{-6}$ for $(\ell,b)=(331.09^\circ,-2.42^\circ)$
$\gamma$ Nor; $\langle\tau\rangle=0.72^{+0.41}_{-0.28}\times 10^{-6}$ for $(\ell,b)=(18.51^\circ,-2.09^\circ)$
$\gamma$ Sct; $\langle\tau\rangle=0.30^{+0.23}_{-0.20}\times 10^{-6}$ for $(\ell,b)=(26.60^\circ,-2.15^\circ)$
$\beta$ Sct.
In this case we use $\rho_l\equiv \rho_b+\rho_d$ in equation \eqref{weightedOD} and replace $dn_s/dD_s$ by a gaussian of mean 7 kpc and standard deviation 1 kpc in good agreement with the source distance distribution shown in Figure 11 of Ref.~\cite{2009A&A...500.1027R}.

\end{itemize}
The results of this cross-check are presented in Figure \ref{fig:depth}, where the boxes represent the experimental measurements and corresponding 1$\sigma$ uncertainties, while the blue dots and error bars refer to the theoretical expectations of the five baryonic models once normalised to the MACHO CGR result in equation \eqref{macho}. It is evident that we obtain a very good agreement for all models, since the expected ranges fall nicely within less than $1\sigma$ off the measurements, the only exceptions being models 1 and 2 for the spiral arm $\beta$Sct (but in this case the discrepancy is just slightly above 1$\sigma$ and so we shall not consider it worrying). This is in fact a reassuring result.

\par Notice that we reproduce the experimental results not only in the innermost galactic fields reached by OGLE-II \cite{Sumi:2005da} and EROS \cite{Hamadache:2006fw}, but also beyond the limits of the central galactic regions where EROS collected microlensing events from four spiral arms across the sky \cite{2009A&A...500.1027R}. It is hence fair to state that the five models defined in Section \ref{MWmodel} and normalised to 2005 MACHO optical depth in equation \eqref{macho} are in good agreement with present microlensing data. Moreover, we have further checked that the corresponding bulge masses are well within the values found in literature $M_b\simeq 1-2\times10^{10}\textrm{ M}_\odot$ \cite{Dwek:1995xu,Klypin,Dehnen,Novatin:2007dd}. We shall use these five models in the remainder of the present work.

\par Several comments are in order here. First, let us stress that all microlensing data used up to now (including the 2005 MACHO result \cite{Popowski:2004uv}, which is the one adopted throughout our analysis) are sensitive to Einstein times 3 $\lesssim t_E/\textrm{days} \lesssim $ 700, or in lens masses $1.4\times 10^{-3} \lesssim M_l/\textrm{M}_\odot\lesssim 79$ (in this simple estimate we have assumed the typical values $v_T=100$ km/s, $D_s=10$ kpc and $D_l=0.5D_s$). This means that compact, baryonic objects of virtually all masses (including all stars and the most massive planets) contribute to the observed optical depth to which we are normalising our bulge models. 

\par In particular, less massive planets ($M_l/\textrm{M}_\odot \lesssim 1.4\times 10^{-3}$) could also contribute significantly to the galactic mass budget, and one might wonder whether the population of unbound Jupiter-mass objects recently discovered through microlensing \cite{Sumi:2011kj} could lead to a systematic underestimate of the baryonic model normalisation. The presence of these objects has been invoked to explain MOA and OGLE microlensing events with $t_E<2$ days, and their abundance has been estimated to be approximately twice that of main-sequence stars \cite{Sumi:2011kj}. Their contribution to the mass density in the bulge/bar and disk is 
\begin{equation}
f_{\textrm{PL}}=\frac{\int { dM \, M (dN/dM)_{\textrm{PL}} }}{\int { dM \, M (dN/dM)_{\textrm{stars}} }} \quad ,
\end{equation}
where $(dN/dM)_{\textrm{PL}}$ and $(dN/dM)_{\textrm{stars}}$ are the mass functions of the Jupiter-mass population and the stars, respectively. Taking a Dirac delta for $(dN/dM)_{\textrm{PL}}$ and model 1 of Table S3 of the supplementary information of Ref.~\cite{Sumi:2011kj}, we find $f_{\textrm{PL}}\simeq 0.1\%$. We shall therefore neglect this newly discovered population in our calculations, since its contribution is much smaller than the uncertainties in the mass model.

\begin{figure*}[htp]
\centering
\includegraphics[width=0.4\linewidth]{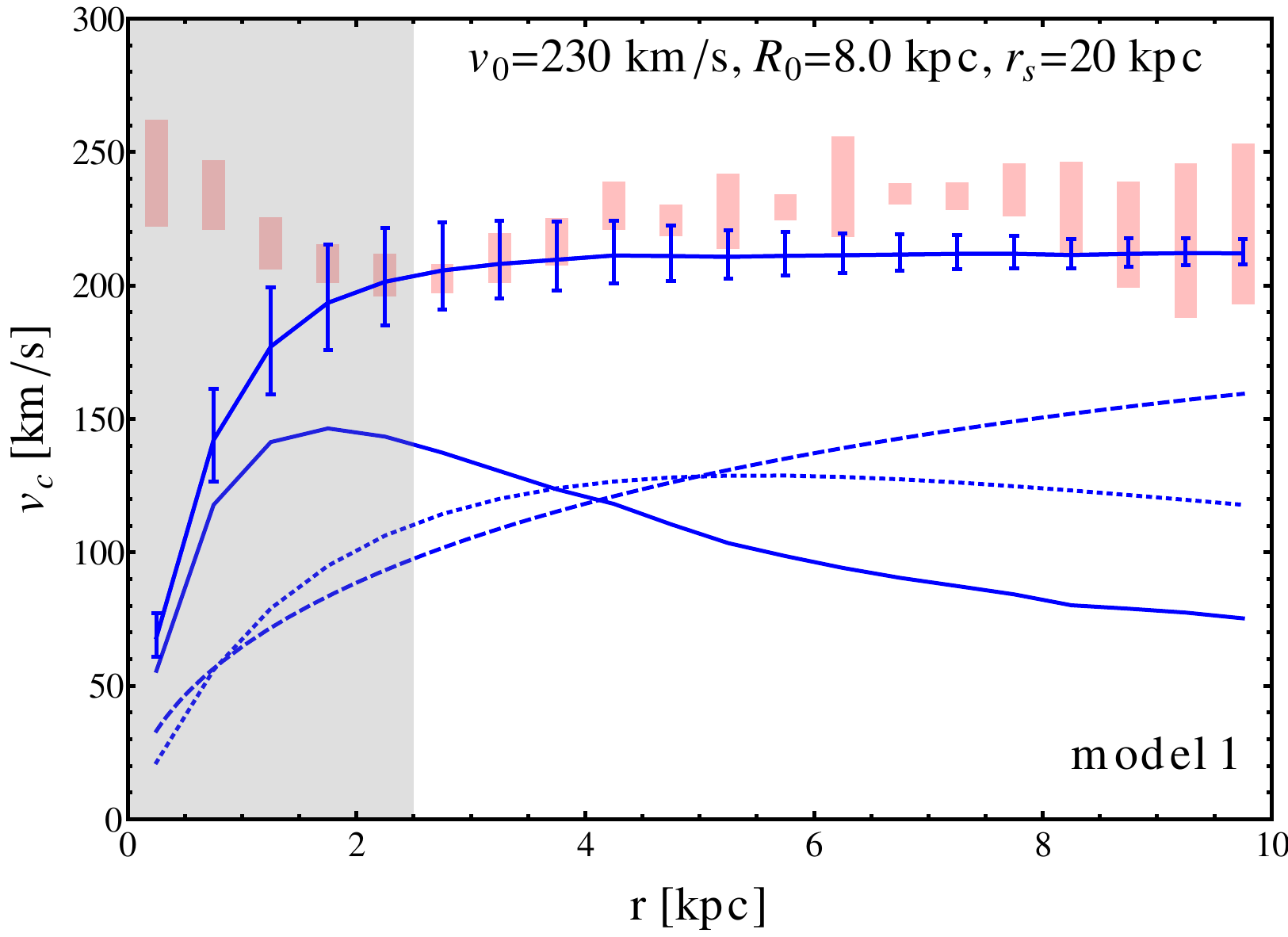}
\includegraphics[width=0.4\linewidth]{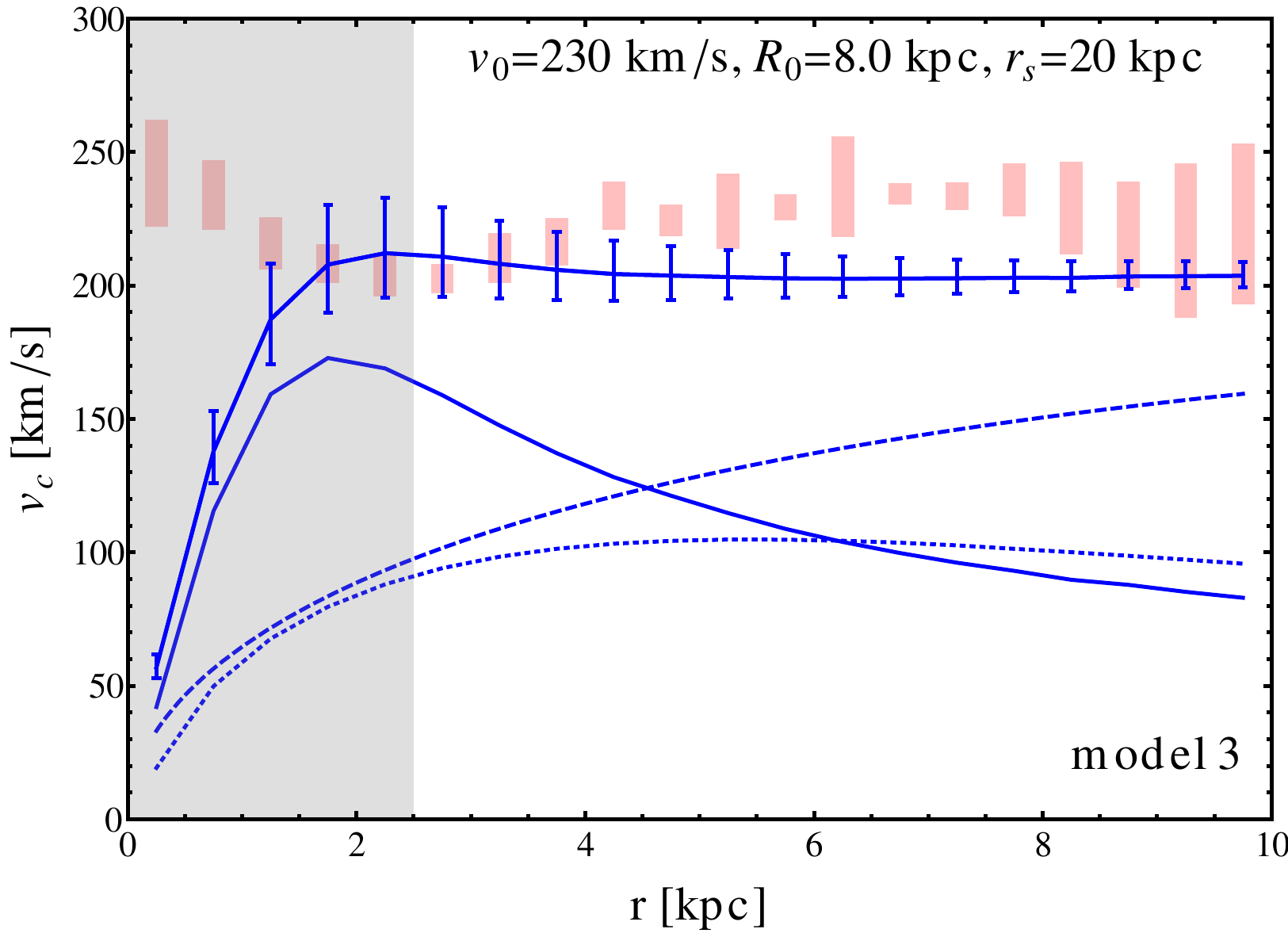}
\includegraphics[width=0.4\linewidth]{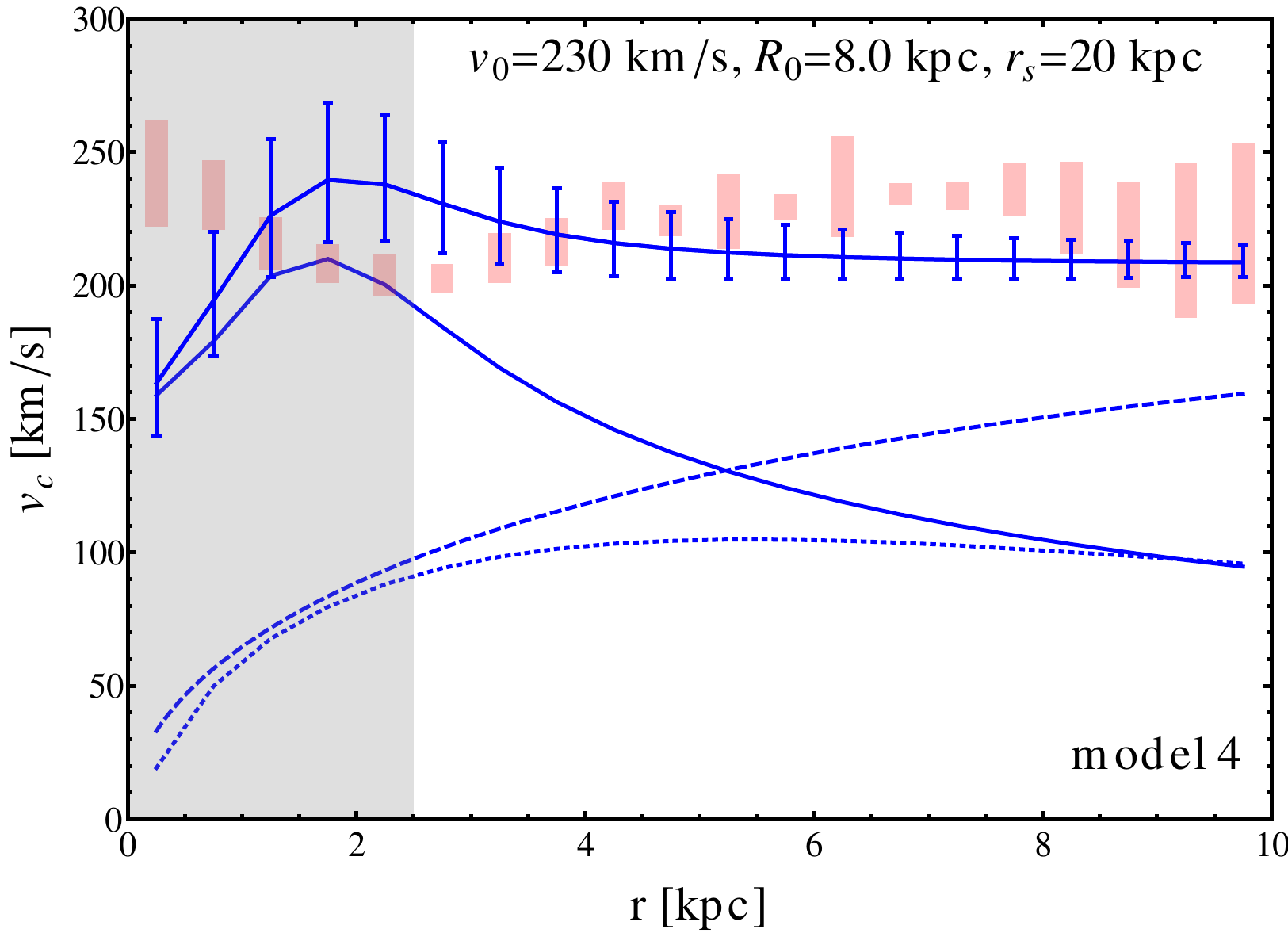}
\includegraphics[width=0.4\linewidth]{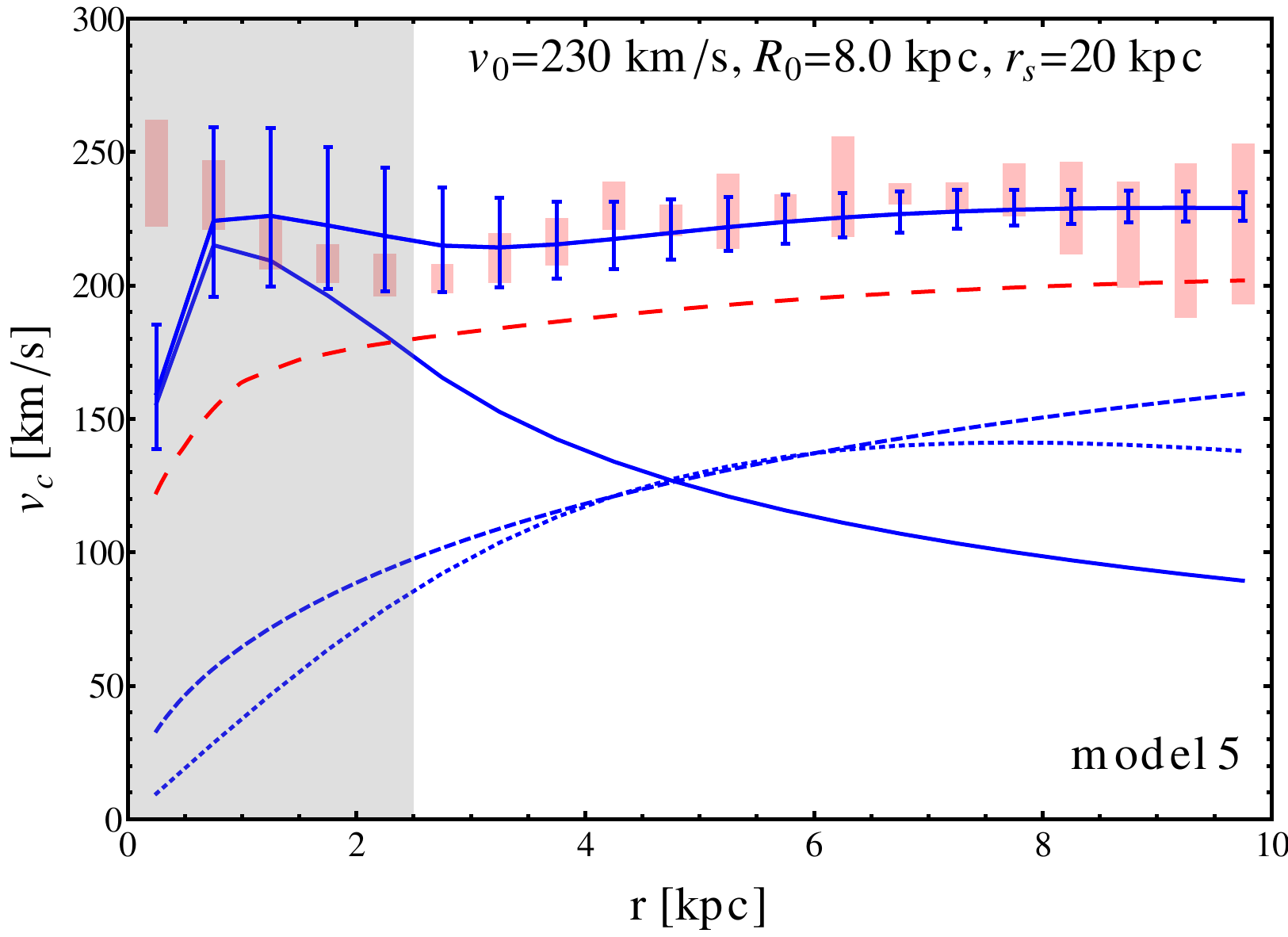}
\caption{The break-down of the rotation curve $v_c(r)$ for different Milky Way mass models. The lower solid, dotted and dashed lines correspond to the individual contributions of bulge/bar, disk and halo, respectively. In all frames the adopted Dark Matter halo follows a spherical NFW profile with $r_s=20$ kpc, $\alpha=1$ and $\rho_0=0.4 \textrm{ GeV/cm}^3$. The rotation curve for model 2 has been omitted since it is very similar to the one featured in model 1. The upper blue solid line with error bars indicates the expected total rotation curve, while the red (dark) boxes show the compilation of data in \cite{Sofue:2008wt} rescaled to $R_0=8$ kpc and $v_0=230$ km/s. The leftmost shaded area shows the cut $r\geq2.5$ kpc used throughout the analysis. The red long-dashed line in the bottom right frame shows the adiabatic compression of an initial NFW profile with $r_s=20$ kpc, $\alpha=1$ and $\rho_0=0.4 \textrm{ GeV/cm}^3$ and assuming model 5 for the baryonic mass distribution. In these plots all uncertainties are 1$\sigma$.} 
\label{fig:vc}  
\end{figure*}

\section{Rotation curve}\label{rotcurve}
\par This Section is dedicated to outlining the formalism and observational status regarding one key dynamical observable, the galactic rotation curve.

\subsection{Theoretical framework}\label{rotcurvetheo}
\par Regardless of their nature, all forms of {\it matter} contribute to the rotation curve of our Galaxy, unlike the case of microlensing, where only compact bodies along the line of sight play a role. This is basically what allows us to extract information about the DM and gas components. In full generality, the circular velocity $v_c$ at a given galactocentric distance $r$ reads
\begin{equation}\label{vc}
v_c^2(r) = \sum_i{v_{c,i}^2(r)} = \sum_i{r \, \frac{d\phi_i}{dr}\left(r,\theta=\pi/2,\varphi\right)   } \quad , 
\end{equation}
where $i$ runs over all the mass components (bulge/bar, disk and Dark Matter halo), $\phi_i$ is the gravitational potential associated to component $i$ and $(r,\theta,\varphi)$ are galactic spherical coordinates ($\theta=\pi/2$ defines the plane of the galactic disk). For the spherical Dark Matter halos in equations \eqref{NFW} and \eqref{Ein} one recovers the well-known expression $v_{c,DM}^2(r)=G M_{DM}(<r)/r$, where $M_{DM}(<r)$ is the DM mass enclosed in a sphere of radius $r$ around the galactic centre. The case of an oblate halo referred to at the end of Section \ref{MWmodel} can also be treated analytically with a slightly more complicated expression (cf.~equation (2-91) in \cite{BinneyTremaine}). As for the baryonic contribution, let us notice that all components of model 5 are specified through their gravitational potential, which renders trivial the computation of the corresponding circular velocity with equation \eqref{vc}. Finally, since the mass distributions of models 1--4 are rather complicated and triaxial in general, there is no simple expression for $v_c$; in this case we compute the gravitational potential due to an arbitrary mass distribution by expressing the solution of the Poisson equation as a series of multipoles up to order $l_{max}=2$ (cf.~equation (2-122) in \cite{BinneyTremaine}) and then apply equation \eqref{vc}. All theoretical expectations for $v_c$ presented in the following have been averaged over $\varphi$ in order to ease comparison with experimental data (which refer to different positions across the galactic plane). 

\par With this formalism at hand we can finally compute the rotation curve associated to the different models specified in Section \ref{MWmodel} and whose bulge/bar components were appropriately normalised to microlensing data (see Section \ref{microl} and Figure \ref{fig:depth} for details). Figure \ref{fig:vc} displays the rotation curve break-down for our baryonic models (models 1 and 2 yield very similar rotation curves despite the different bulge shapes, so we omit the latter model for plotting purposes) and selecting a spherical NFW profile with $r_s=20$ kpc, $\alpha=1$ and $\rho_0=0.4 \textrm{ GeV/cm}^3$. The bulge/bar, disk and halo contributions are represented by the lower solid, dotted and dashed curves, respectively, in the range $r=0-10$ kpc with bins of 0.5 kpc. Notice that for model 5 the disk contribution in dotted contains both the stellar and gas components. The upper blue solid curve with error bars in every frame of Figure \ref{fig:vc} denotes the total rotation curve predicted by each model. The error bars show the propagated uncertainty due to the bulge normalisation scaled to MACHO microlensing optical depth in equation \eqref{macho}. As we shall see, the combination of microlensing data and the rotation curve sets an upper limit on the distribution of Dark Matter across the Galaxy. Before deriving these constraints, we turn to the status of present data on the galactic rotation curve, represented by the red (dark) boxes in Figure \ref{fig:vc}.

\subsection{Observations}\label{secrotObs}

\par Different methods are available to infer the galactic rotation curve at galactocentric distances $r\lesssim 10$ kpc, which is the range we are interested in for the scope of this paper. One of the most used techniques relies on the observation of gas clouds moving in the galactic plane. For each line of sight $(\ell,b)$ the extreme velocity $v_t$ -- the so-called ``terminal velocity'' -- is measured and converted to the circular velocity at a given radius under the assumption of strictly circular gas orbits \cite{Malhotra}: $v_c(R_0\sin \ell)=v_t(\ell)+v_0 \sin \ell$, where $v_0\equiv v_c(R_0)$ is the local circular velocity. The 21 cm line is widely used in literature to determine the terminal velocity of gas clouds in the inner regions of the Milky Way. Other kinematical probes for the outer regions of our Galaxy  are C-stars, observed through near-infrared photometry (e.g.~\cite{DemersBattinelli}), and the CO emission line of HII regions (e.g.~\cite{Blitzetal}).

\par Generically, the observed velocities of either gas clouds or stars with respect to the Earth must be transformed to a circular velocity through specific assumptions on the local reference frame. The different values adopted by the several observational groups over the years make it difficult to simply take their final values altogether. In Ref.~\cite{Sofue:2008wt} the authors construct a rotation curve of the Galaxy starting from the very observational data, and unify the inference of rotation curves for a single set of local galactocentric radius and local circular velocity $(R_0,v_0)=(8.0\textrm{ kpc},200\textrm{ km/s})$. Here we use that compilation of data, which includes measurements from several observational groups and techniques. 

\par The values of $(R_0,v_0)$ adopted in the literature vary significantly, and since the constraints we shall obtain in Section \ref{results} depend quite sensitively on the choice of the Earth reference frame, we will take into account the error on $R_0$ and $v_0$ throughout the analysis in order to ``bracket'' the uncertainty on the constraints. In addition to the range $R_0=8.0\pm 0.5$ kpc mentioned in Section \ref{MWmodel}, we consider local circular velocities in the interval $v_0=230\pm 30$ km/s. This choice reasonably encompasses a broad set of measurements \cite{Xue2008,Sofue2008,Reid2009,Bovy2009,McMillan2009}. Recent analyses (e.g.~\cite{Reid2009}) in particular seem to indicate local velocities well in excess of the 1985 IAU recommended value $v_0=220$ km/s \cite{IAU1985b}. In order to convert the data in \cite{Sofue:2008wt} to a given set $(R'_0,v'_0)$, the following transformation laws are applied:
\begin{equation}\label{scaling}
R'= R\frac{R'_0}{R_0} \quad ; \quad  \quad v'_t= v_t +\frac{R}{R_0}(v'_0-v_0)
\end{equation}
with the {\it primed} quantities being the ones obtained with our new choice of $(R'_0,V'_0)$, and the {\it plain} ones being those as from \cite{Sofue:2008wt}, with $(R_0,v_0)=(8.0\textrm{ kpc},200\textrm{ km/s})$. These transformation laws are only strictly valid for the measurement of terminal velocities. However, since our constraints will come essentially from radii within the Solar circle, where terminal velocity data points dominate the compilation in Ref.~\cite{Sofue:2008wt}, this procedure is reasonable.

\par Given that we are using a compilation of data obtained with different techniques, the treatment of uncertainties on the derived quantity $v_c(r)$ is a delicate matter. The following strategy is adopted in this work. After scaling the data from \cite{Sofue:2008wt} to a given $(R_0,v_0)$ according to equations \eqref{scaling}, a binning of 0.5 kpc is applied in the galactocentric distance range $r=0-10$ kpc. Since several measurements lie in each $r$--bin, the central value on $v_c$ is taken to be the mean of all the central values in the bin, while the corresponding $1\sigma$ systematic uncertainty is the standard deviation of the central values. We estimate the $1\sigma$ statistical uncertainty as the mean of the $1\sigma$ uncertainties of all the measurements in each bin, and add it to the systematic error. It is this total uncertainty that is represented by the red (dark) boxes in Figure \ref{fig:vc} (for $R_0=8$ kpc, $v_0=230$ km/s) and that we use to gauge our knowledge on the Milky Way rotation curve in the following Section. 

\par We restrict our analysis to galactic radii $r\geq2.5$ kpc, corresponding to longitudes $|\sin \ell|\geq 0.30-0.35$ in line with what is usually done in the literature \cite{Dehnen,CatenaUllio}. This cut -- marked in the plots of Figure \ref{fig:vc} -- is imposed by the severe triaxiality of the gravitational potential in the inner galactic regions caused by the presence of the galactic bar, invalidating the assumption of circular orbits.

\begin{figure*}[htp]
\centering
\includegraphics[width=0.45\textwidth]{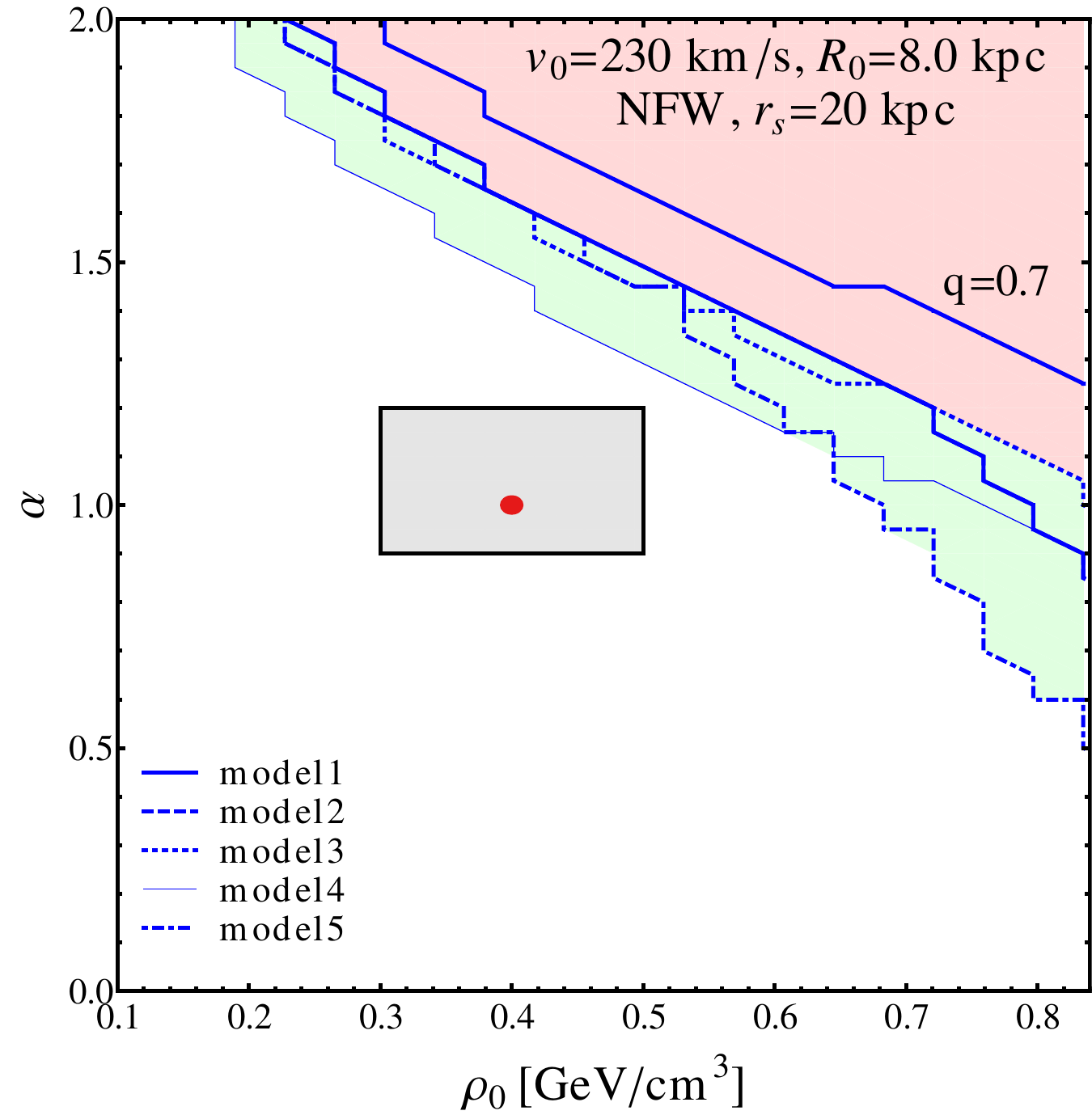}
\includegraphics[width=0.45\textwidth]{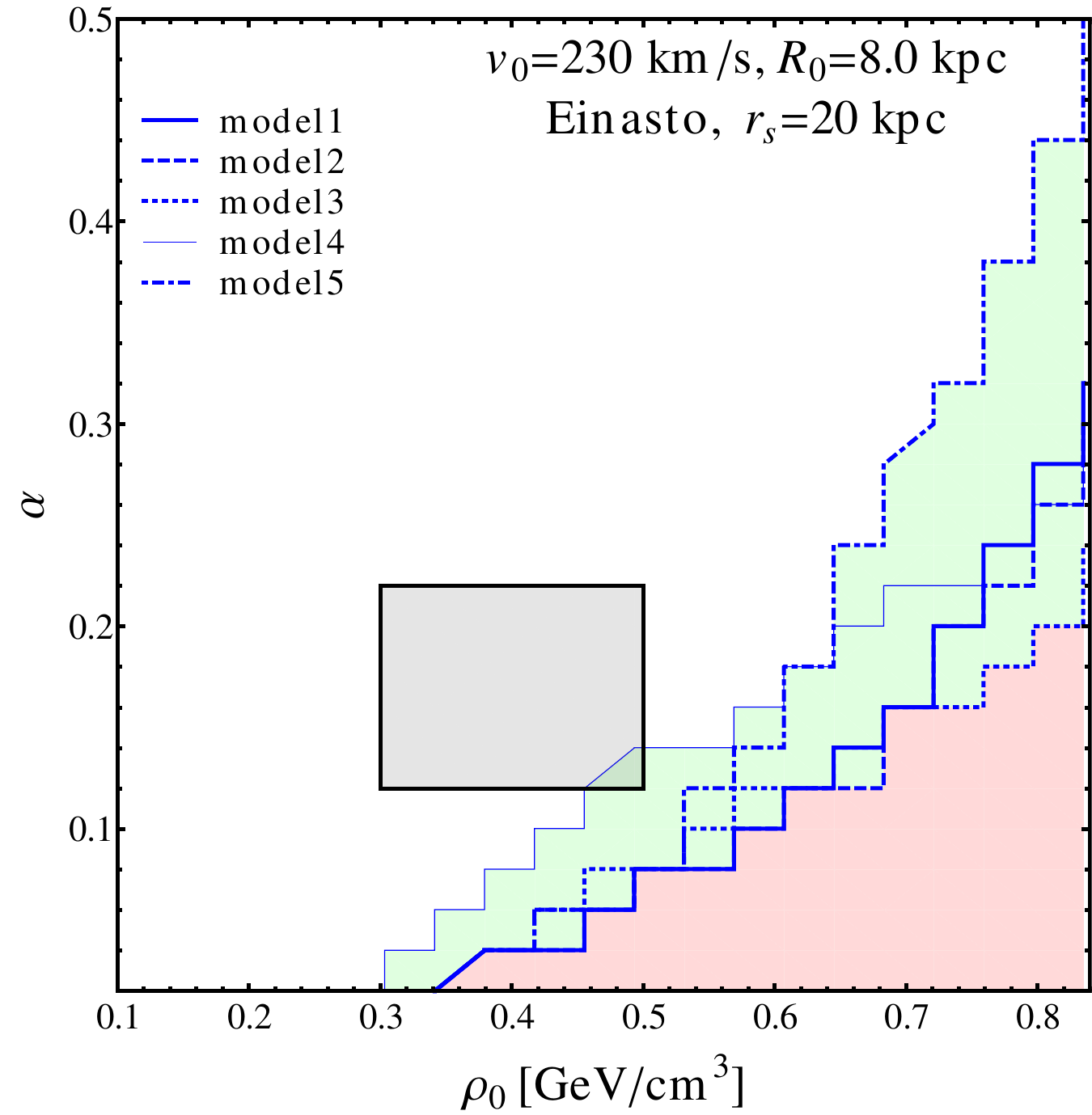}
\caption{Constraints on the Dark Matter distribution parameters $\rho_0$ and $\alpha$ provided by current data on the Milky Way rotation curve for a generalised NFW (left) and an Einasto (right) profile. The thick solid, thick dashed, thick dotted, thin solid and thick dot-dashed lines are the 2$\sigma$ constraints in the case where the galactic baryonic component follows model 1, 2, 3, 4 and 5, respectively, all scaled to match the microlensing optical depth in equation \eqref{macho}. The green (light) shadowed area delimits the uncertainty on the constraint given the present-day baryonic models, while the red (dark) shadowed region indicates the excluded parameters. In the left frame, the upper thick solid line labelled ``$q=0.7$'' refers to the constraints for an oblate NFW profile with axis ratio $q=0.7$ in the case of taking model 1 for the baryonic component. The shadowed rectangle encompasses the ranges of profile slopes found in numerical simulations and the values of $\rho_0$ found in the recent literature (see Section \ref{MWmodel}), while the red filled circle in the left frame marks the parameter set $(\rho_0=0.4\textrm{ GeV/cm}^3,\alpha=1.0)$ used to produce Figure \ref{fig:vc}. In both frames we have fixed $r_s=20$ kpc, $R_0=8.0$ kpc and $v_0=230$ km/s.
}
\label{fig:nfwein}  
\end{figure*}

\section{Results}\label{results}

\par We now turn to the discussion of our results. From Figure \ref{fig:vc} one can see how the modelling of the baryonic component in our Galaxy affects the rotation curve. First of all, let us recall that the expected circular velocity depicted in that Figure is obtained by tuning our bulge models to microlensing data, independently of the data on the rotation curve itself. It is thus remarkable that in all frames the expectation roughly matches the data -- this argues in favour of the accurateness of present Milky Way mass models in explaining both microlensing and dynamical observables. Models 1--4, in particular, succeed in reproducing a flat rotation curve for $r\gtrsim 2.5$ kpc but appear slightly below the data at $6\textrm{ kpc}\lesssim r \lesssim 10$ kpc, while model 5 does feature the exact same pattern indicated by the data. Again, we shall not regard the inner 2.5 kpc to set our constraints. 

\par Notice as well that the coarse $r-$binning may be misleading at small scales; model 5, for instance, produces a steep bump in $v_c$ at $r\lesssim 0.5$ kpc (see e.g.~Ref.~\cite{Gardner:2010pd}) which is not seen in Figure \ref{fig:vc} simply because of the chosen bins.

\subsection{Conservative constraints on $(\rho_0,\alpha)$}

\par Figure \ref{fig:vc} assumes an NFW profile with fiducial values $\rho_0=0.4\textrm{ GeV/cm}^3$, $\alpha=1.0$ (and $r_s=20$ kpc). We are now interested in studying the constraints that can be set in the DM parameter space $(\rho_0,\alpha)$ by using the data on the Milky Way rotation curve. A given set $(\rho_0,\alpha)$ will be considered to be excluded at $2\sigma$ if the $2\sigma$ lower end of the expected $v_c$ (curve with error bars in Figure \ref{fig:vc}) exceeds the $2\sigma$ upper end of the measured $v_c$ (boxes in Figure \ref{fig:vc}) in at least one radial bin in the range $2.5\textrm{ kpc}\leq r \leq 10$ kpc. This is a rather conservative procedure indeed -- note, for instance, that for model 4 the fiducial NFW setup (see bottom left frame of Figure \ref{fig:vc}) is not excluded at $2\sigma$, even though the predicted rotation curve is far from explaining the data.

\par The constraints obtained with these definitions are shown in Figure \ref{fig:nfwein} for the different models 1--5 and the generalised NFW (left) and Einasto (right) DM profiles. Red (dark) shading signals the excluded regions and green (light) shading spans the uncertainty on the exact constraint given the present-day baryonic models. These plots assume the fiducial setup $r_s=20$ kpc, $R_0=8.0$ kpc, $v_0=230$ km/s. It is worth mentioning that the constraints presented in Figure \ref{fig:nfwein} are generally dominated by two radial bins: the one centred at $r=2.75$ kpc for small local DM densities and the one at $r=7.75$ kpc for larger densities.

\par The conservative constraints shown in Figure \ref{fig:nfwein} are found to be consistent with the range of inner slopes/shape parameters $\alpha$ found in numerical simulations and local DM densities $\rho_0$ hinted by dynamical observables (see Section \ref{MWmodel} and shadowed rectangles in Figure \ref{fig:nfwein}). Hence, contrary to the findings of Ref.~\cite{BinneyEvans}, we conclude that microlensing observations and data on the rotation curve of the Milky Way are perfectly compatible with the findings of $N-$body simulations in $\Lambda$CDM cosmologies. Nonetheless, interesting constraints may be placed on the slope of the DM profile in our Galaxy: for the fiducial density $\rho_0=0.4\textrm{ GeV/cm}^3$, Figure \ref{fig:nfwein} excludes $\alpha\gtrsim 1.5$ ($\alpha\lesssim 0.06$) for the generalised NFW (Einasto) profile. Notice that in the NFW (Einasto) case one can set an upper (a lower) limit on $\alpha$ -- this is simply because for $r\ll r_s$, $\partial\log\rho_{DM}/\partial\log r=-\alpha$ for NFW and $\partial\log\rho_{DM}/\partial\log r=-2(r/r_s)^{\alpha}$ for Einasto. Therefore, unlike in the NFW case, a larger $\alpha$ for the Einasto profile corresponds to a faster roll of the slope to 0 as $r\to 0$ and thus a less steep profile.

\par Deviations from spherical symmetry can in principle affect the constraints. By modelling an NFW oblate halo as detailed in Section \ref{MWmodel}, we obtain the exclusion curve labelled ``q=0.7'' in Figure \ref{fig:nfwein} (left) for the case of model 1: at first glance this constraint appears weaker than the spherical one, but it should be noted that an oblate profile corresponds to a higher $\rho_0$ (about 20\% higher according to \cite{paperDMlocal}). Departures from spherical symmetry are therefore not able to weaken significantly our constraints.

\par Up to now we have fixed the scale radius, galactocentric distance and local circular velocity to the respective fiducial values, $r_s=20$ kpc, $R_0=8.0$ kpc, $v_0=230$ km/s. These astrophysical parameters, whose uncertainties are sizeable, affect in distinct ways our calculations. The scale radius $r_s$, for instance, sets the concentration of the DM profile; the smaller $r_s$ the larger the DM contribution to the rotation curve. On the other hand, a smaller $R_0$ shrinks the bulge and the disk leading to an increase in the bulge central density to produce the same optical depth; however, a smaller $R_0$ also leads to a less constraining $v_c$ data set so that overall the larger $R_0$ the more aggressive our DM constraints. The local circular velocity $v_0$, instead, sets essentially the plateau of the rotation curve and thus tighter constraints result for smaller $v_0$.

\begin{figure}[htp]
\centering
\includegraphics[width=0.45\textwidth]{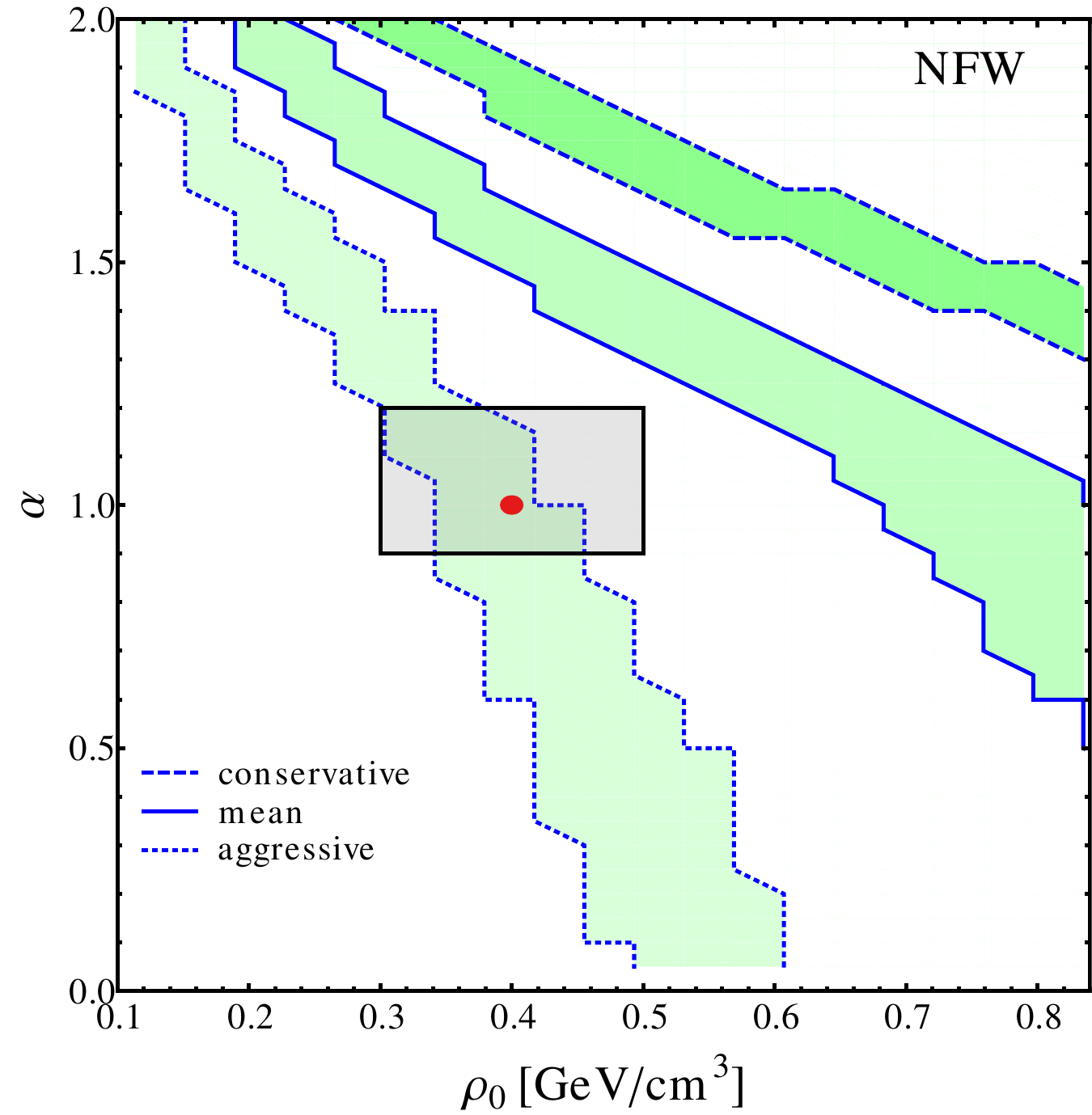}
\caption{The bracketing of the 2$\sigma$ upper limits on the Dark Matter distribution parameters $\rho_0$ and $\alpha$ for the generalised NFW profile and three astrophysical setups: conservative (dashed; $r_s=35$ kpc, $R_0=7.5$ kpc, $v_0=260$ km/s), mean (solid; $r_s=20$ kpc, $R_0=8.0$ kpc, $v_0=230$ km/s) and aggressive (dotted; $r_s=10$ kpc, $R_0=8.5$ kpc, $v_0=200$ km/s). The two lines for each setup encompass the upper limits set using the baryonic models 1--5. In particular, the mean shadowed area as well as the shadowed rectangle are the same as in the left frame of Figure \ref{fig:nfwein}.}
\label{fig:bracket}  
\end{figure}

\par In view of these considerations and using the ranges for $r_s$, $R_0$ and $v_0$ outlined in Sections \ref{MWmodel} and \ref{rotcurve} ($r_s=20^{+15}_{-10}$ kpc, $R_0=8.0\pm0.5$ kpc, $v_0=230\pm30$ km/s), we define three astrophysical setups: \emph{(i)} conservative, with $r_s=35$ kpc, $R_0=7.5$ kpc, $v_0=260$ km/s; \emph{(ii)} mean, with $r_s=20$ kpc, $R_0=8.0$ kpc, $v_0=230$ km/s; and \emph{(iii)} aggressive, with $r_s=10$ kpc, $R_0=8.5$ kpc, $v_0=200$ km/s. The mean configuration was used in Figures \ref{fig:vc} and \ref{fig:nfwein}. Figure \ref{fig:bracket} shows the effect of adopting the conservative or aggressive setups on the derived DM upper limits for the generalised NFW profile. For simplicity we only show the upper limits encompassed by all models, instead of individual constraints. We see from this Figure that, for reasonable local DM densities, an NFW profile in line with the findings of numerical simulations can only be (barely) excluded at the expenses of pushing some astrophysical parameters to somewhat extreme values (in particular $v_0=200$ km/s). We are thus led to the conclusion that the results of Ref.~\cite{BinneyEvans} do not hold, given the available microlensing and dynamical data and our present knowledge on astrophysical parameters such as $r_s$, $R_0$ or $v_0$.

\subsection{Determination of $(\rho_0,\alpha)$}

\par Models 1--4 include no gas component, which in principle makes the corresponding exclusion curves in Figure \ref{fig:nfwein} conservative. Model 5, instead, includes a gas disk and therefore we can go one step further and ask which DM parameters $(\rho_0,\alpha)$ provide {\it the best fit} to the measured rotation curve. 

\begin{figure*}
\centering
\includegraphics[width=0.45\textwidth]{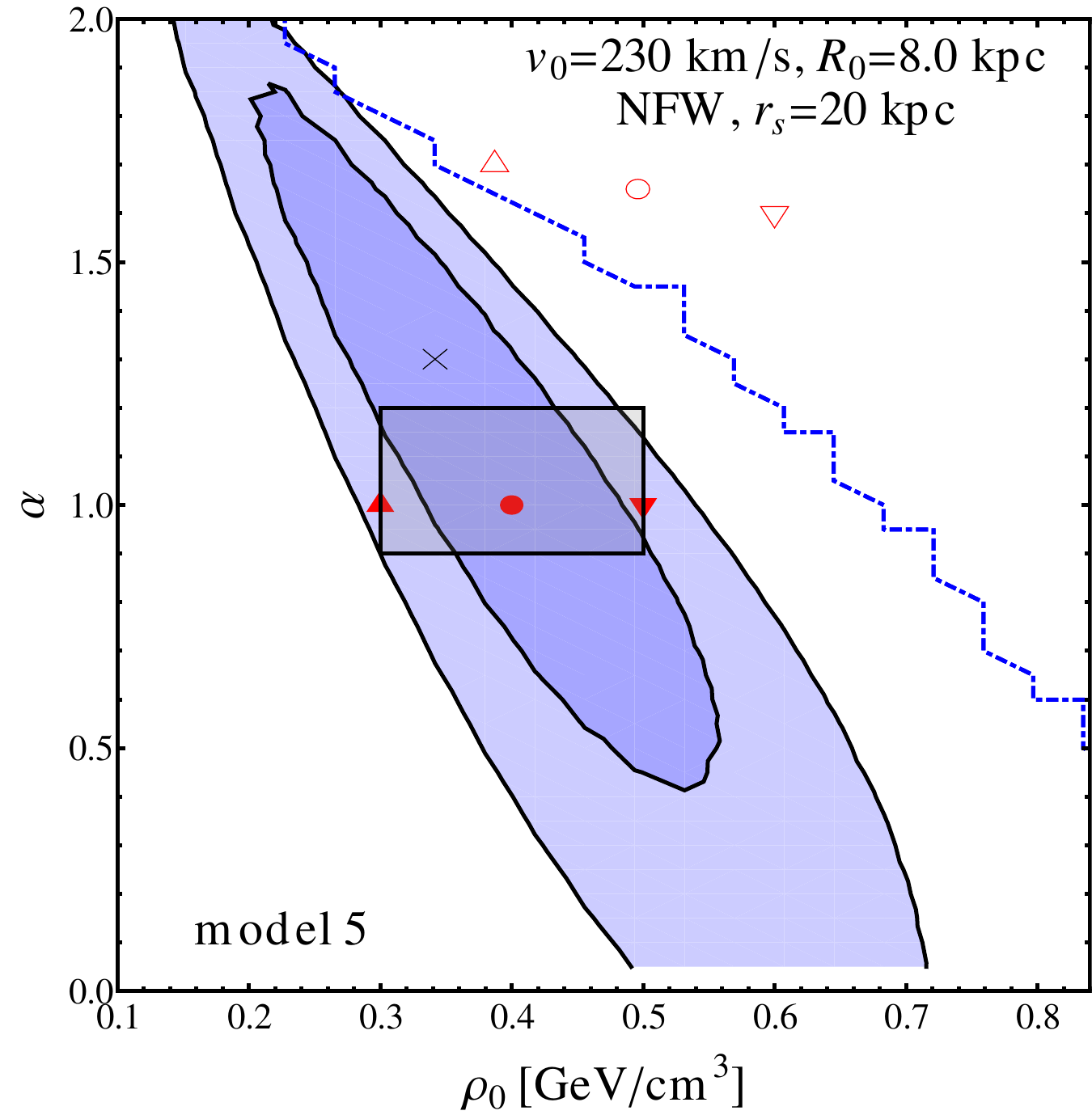}
\includegraphics[width=0.45\textwidth]{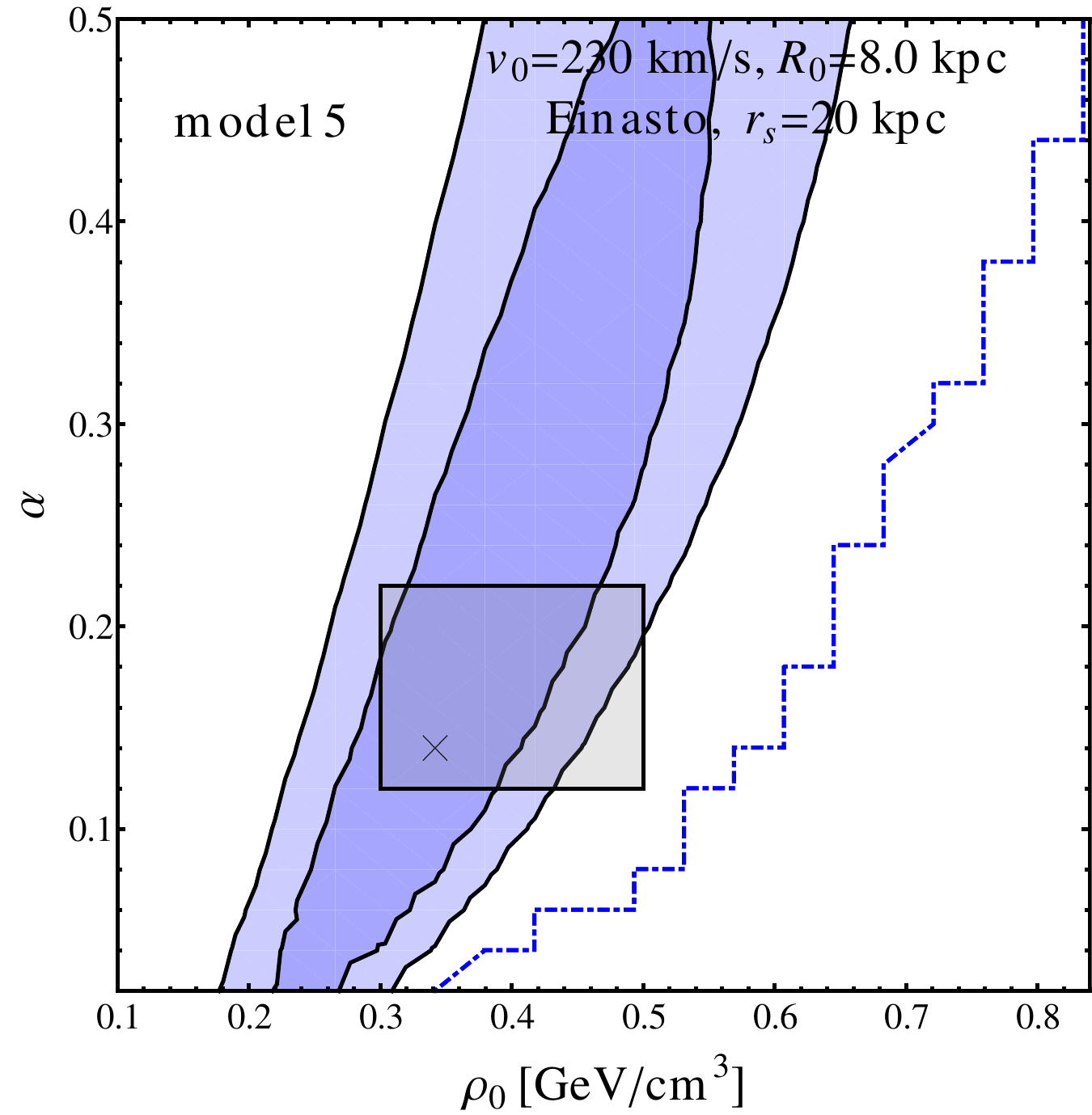}
\caption{Constraints on the Dark Matter distribution parameters $\rho_0$ and $\alpha$ for a generalised NFW (left) and an Einasto (right) profile using the baryonic model 5. The thick dot-dashed curve is the 2$\sigma$ constraint already shown in Figure \ref{fig:nfwein}, while the contours show the parameter space producing a good fit to the rotation curve ($\Delta\chi^2=2.30,6.18$) with the best-fit configuration indicated by the cross. The shadowed rectangle encompasses the ranges of profile slopes found in numerical simulations and the values of $\rho_0$ found in the recent literature (see Section \ref{MWmodel}), while the red filled circle in the left frame marks the parameter set $(\rho_0=0.4\textrm{ GeV/cm}^3,\alpha=1.0)$ used to produce Figure \ref{fig:vc}. The empty up-triangle, circle and down-triangle in the left frame show the local density and shape of the DM profile upon adiabatic contraction of the initial profile indicated by the corresponding filled symbols. The adiabatic contraction was applied using model 5 to fix the baryonic distribution $M_b(<r)$, that entails $f_b=5.2\%,4.0\%,3.0\%$ for the up-triangle, circle and down-triangle, respectively. In both frames we have fixed $r_s=20$ kpc, $R_0=8.0$ kpc and $v_0=230$ km/s.
}
\label{fig:fit}  
\end{figure*}

\par For the fitting procedure we bin the $v_c$ data as discussed in Section \ref{rotcurvetheo} (0.5 kpc bins) and consider the radial range $r=2.5-10$ kpc. The microlensing uncertainties were propagated through the baryonic model adopted, and the sum of statistical and systematic was used for the rotation curve data according to the procedure described in Section \ref{secrotObs}. We have assumed that different bins are uncorrelated: for the rotation curve observations, different datasets dominate the measure (and the uncertainty) at different bins. While this reduces the correlation between different bins, it is also to be taken into account that the adoption of a systematic in addition to statistical is in line with our conservative approach.

\par The results of this fitting procedure are shown by the contours (corresponding to $\Delta \chi^2=2.30,\,6.18$ for a two-parameters fit) in Figure \ref{fig:fit}: for DM parameters inside the contours, model 5 manages to explain both the microlensing optical depth towards the bulge and the rotation curve of our Galaxy. It is interesting (and reassuring) that the contours fall nicely on top of the expected ranges indicated by the shadowed rectangles. In particular, we find 
\begin{eqnarray}\label{rho0NFW}
\rho_0&=& 0.20-0.56\textrm{ GeV/cm}^{3} \quad \text{(NFW)} \\ 
\rho_0&=& 0.22-0.55\textrm{ GeV/cm}^{3} \quad \text{(Einasto)}
\end{eqnarray}
at $1\sigma$, which is consistent with previous estimates obtained with different techniques \cite{CatenaUllio,SaluccilocalDM}.

\par We have checked the robustness of the result in equation \eqref{rho0NFW} in a number of different ways. First, the role of the number and width of the radial bins in the fitting of the rotation curve was analysed. Increasing the size of the bins to 1 kpc leads to a measurement $\rho_0=0.16-0.65\textrm{ GeV/cm}^{3}$ in the case of a generalised NFW profile, i.e.~an increase in the uncertainty on $\rho_0$ from 47\% to 60\% with respect to equation \eqref{rho0NFW}. A smaller binning would lead to a smaller uncertainty on $\rho_0$, but the average uncertainty on the galactocentric position of the rotation curve trackers in Ref.~\cite{Sofue:2008wt} is $\sim$ 0.5 kpc, which therefore provides a physical lower limit on the size of the bins.

Second, as pointed out in Section \ref{secrotObs}, the transformations \eqref{scaling} are only strictly applicable for terminal velocities, which dominate the compilation of Ref.~\cite{Sofue:2008wt} for $r<R_0$. In the range $R_0\leq r\leq 10$ kpc, the dominant data points on $v_c(r)$ are due to the measurement of velocities in CO and HII regions. In that case the scaling for a given set $(R'_0,v'_0)$ is different from \eqref{scaling} and reads $v'=v+(v'_0-v_0)$ if $R'_0=R_0=8$ kpc. This means that up to now we have been overestimating the scaled velocities in the range $8\textrm{ kpc}\leq r \leq10$ kpc by $\lesssim 3$\% for $v'_0=230$ km/s. We have however checked that applying $v'=v+(v'_0-v_0)$ for $8\textrm{ kpc}\leq r \leq10$ kpc and the set of transformations \eqref{scaling} for $r<8$ kpc changes the 1$\sigma$ determination of $\rho_0$ by a small (but non-negligible) amount: $\rho_0=0.17-0.50\textrm{ GeV/cm}^{3}$ in the case of NFW profiles.

\par It is also interesting to see how the determination \eqref{rho0NFW} for NFW profiles changes for oblate halos. Using the axis ratio $q=0.7$, we obtain the 1$\sigma$ range $\rho_0=0.26-0.68\textrm{ GeV/cm}^{3}$. Using instead a more spherical halo with $q=0.9$, our 1$\sigma$ local DM density determination reads $\rho_0=0.22-0.59\textrm{ GeV/cm}^{3}$, to compare with the recent measurement based on the vertical motion of stars in the solar neighbourhood \cite{Garbari:2011dh,Garbari:2011wi}. Let us note that the up-shifting of the central value (with respect to the spherical case) is expected since the oblateness allows for the concentration of more DM along the galactic plane for an assigned halo mass.

\par We have also tested the adiabatic contraction model of Ref.~\cite{Blumenthal} (outlined in Section \ref{MWmodel} and equation \eqref{AC}) in the case of the NFW profile. Using once again model 5 for the baryonic component, we have contracted the initial profiles indicated in Figure \ref{fig:fit} (left) by the filled up-triangle, circle and down-triangle with $f_b=M_b(<200\textrm{ kpc})/M_{tot}(<200\textrm{ kpc})=5.2\%,4.0\%,3.0\%$, respectively. The final DM profile turns out to be well fitted by a generalised NFW function with parameters marked by the empty symbols in the same Figure (the contracted profile corresponding to the filled circle is indicated by the red long-dashed line in the bottom right frame of Figure \ref{fig:vc}). In particular, we find enhanced local DM densities and slopes $\alpha\simeq1.6-1.7$, which are slightly above the value $\alpha=1.5$ found elsewhere \cite{BertoneMerritt} (see also references therein) but note that we are using the original adiabatic contraction model \cite{Blumenthal} and not one of its refinements \cite{Gnedin2004,Gustafsson2006}. Although our analysis cannot rule out the presence of adiabatically compressed profiles since they depend on the initial total mass distribution and on the specific baryonic model adopted, it definitely allows us to claim that if the present-day DM profile is steeply rising towards the centre, then the local DM density must be small. For the specific case of $\alpha =1.5$ $(1.7)$ we find an 1$\sigma$ range $\rho_0\simeq0.25-0.34$ $(0.22-0.28)$ GeV/cm$^3$. Some of the extreme models discussed in the literature, e.g. in the context of indirect DM searches \cite{BertoneMerritt,Bertone:2005xv}, are therefore found to be ruled out by a combination of microlensing and dynamical observations.

\section{Conclusions}

\par We have studied the constraints that microlensing and dynamical observations can set on the distribution of Dark Matter in the Galaxy, keeping into account all experimental uncertainties. Starting from state-of-the-art models for the galactic baryonic component, we have rescaled them to match the observed microlensing optical depth towards the galactic bulge, and compared the resulting rotation curve with the one inferred from terminal velocities of gas clouds and other kinematical probes. 

\par This allowed us to revisit the compatibility of different observational probes with the results that emerge from numerical simulations in $\Lambda$CDM cosmologies. We have followed two different approaches. In the first one, we have set conservative upper limits on the Dark Matter local density and profile shape towards the centre of the Galaxy, working with generalised NFW and Einasto profiles. The fiducial parameters usually adopted in the literature for both profiles have been found to be safely within the allowed regions set by our analysis, contrary to earlier claims of inconsistency between observations and cuspy Dark Matter profiles. 

\par In our second approach, we focussed on the only baryonic model among those discussed here that also contains a description of the amount and distribution of gas, which is expected to provide a non-negligible contribution to the mass and therefore to the rotation curve of the Milky Way. For this specific model, we were able to calculate the values of the Dark Matter parameters that provide the best fit to the combination of microlensing and rotation curve data. The resulting 1$\sigma$ range for the local DM density, for NFW profiles, was found to be $\rho_0=0.20-0.56\textrm{ GeV/cm}^{3}$, therefore consistent with estimates obtained in the literature with different techniques. 

\par Finally, we have studied the consequences of adiabatic compression, often invoked as a mechanism that could increase the amount of Dark Matter in the innermost regions of the Galaxy, and found that although our analysis is not able to discard this mechanism in general, it rules out combinations of local densities and profile slopes invoked in the literature, e.g.~in the context of indirect Dark Matter searches. For NFW profiles with $\alpha =1.5$ $(1.7)$ we constrain the local DM density to be within the 1$\sigma$ range $\rho_0\simeq0.25-0.34$ $(0.22-0.28)$ GeV/cm$^3$.

\par As numerical simulations that include the effect of baryons become ever more accurate and reliable, it will be interesting to compare the findings of these simulations with the wealth of observational data that are currently available. In particular, these simulations should allow us to tie together the distribution of baryons and Dark Matter, and to provide a more precise prescription for the slope of the DM profile towards the galactic centre. This slope is currently extrapolated from simulations {\it without} baryons, which is an unreliable procedure given that baryons are known to dominate the gravitational potential in the inner Galaxy. 

\vspace{0.5cm}
\par {\it Acknowledgements:} We would like to thank Sebastiano Calchi Novati, Victor P.~Debattista, Juerg Diemand, Silvia Garbari, Esko Gardner, Michael Kuhlen and Justin Read for helpful discussions. F.I.~ is supported from the European Community research program FP7/2007/2013 within the framework of convention \#235878, and acknowledges the hospitality of the Institute for Theoretical Physics at the University of Z\"urich. M.P.~acknowledges the support from Funda\c{c}\~{a}o para a Ci\^encia e Tecnologia (Portuguese Ministry of Science, Technology and Higher Education) under the program POPH
co-financed by the European Social Fund in the early stages of this work, and from the Swiss National Science Foundation in the later ones.


\bibliographystyle{h-elsevier}
\bibliography{microlensing}

\end{document}